\documentclass[pdflatex,sn-mathphys-num]{sn-jnl}% Math and Physical Sciences Numbered Reference Style
\usepackage{graphicx}%
\usepackage{multirow}%
\usepackage{amsmath,amssymb,amsfonts}%
\usepackage{amsthm}%
\usepackage{mathrsfs}%
\usepackage[title]{appendix}%
\usepackage{xcolor}%
\usepackage{textcomp}%
\usepackage{manyfoot}%
\usepackage{booktabs}%
\usepackage{algorithm}%
\usepackage{pdflscape}
\usepackage{svg}%
\usepackage{algorithmicx}%
\usepackage{algpseudocode}%
\usepackage{listings}%
\usepackage{placeins}%
\theoremstyle{thmstyleone}%
\theoremstyle{thmstyletwo}%

\theoremstyle{thmstylethree}%

\begin{document}

\title[Article Title]{A Voltage-controlled MTJ–CMOS Neuron Emulating Tunable Izhikevich‑Inspired Dynamics
}

%Voltage-controlled MTJ and CMOS Enabled Izhikevich-inspired Reconfigurable Neurons

%Izhikevich-inspired Reconfigurable Neuron using Voltage-Controlled Magnetic Tunnel Junctions Co-Designed with CMOS 


\author[1]{\fnm{Kayode Oluwaseyi} \sur{Adebunmi}}\email{adebunmi@wisc.edu}

\author[2]{\fnm{Jordan} \sur{Athas}}\email{jordan.athas@northwestern.edu}

\author[2]{\fnm{Allison } \sur{Fleming}}\email{allisonfleming2030@u.northwestern.edu}

\author[3]{\fnm{Hamed} \sur{Poursiami}}\email{hpoursia@gmu.edu}

\author[3]{\fnm{
Giorgio A.} \sur{Ascoli}}\email{ascoli@gmu.edu}

\author[3]{\fnm{Maryam} \sur{Parsa}}\email{mparsa@gmu.edu}

\author[2]{\fnm{Pedram Khalili} \sur{ Amiri}}\email{pedram@northwestern.edu}

\author*[1]{\fnm{ Akhilesh R} \sur{Jaiswal}}\email{akhilesh.jaiswal@wisc.edu}

\affil*[1]{\orgdiv{Electrical and Computer Engineering}, \orgname{University of Wisconsin–Madison, Madison}, \orgaddress{\street{ 500 Lincoln Dr}, \city{Madison}, \postcode{53706}, \state{Wisconsin}, \country{USA}}}

\affil[2]{\orgdiv{Electrical and Computer Engineering}, \orgname{Northwestern University}, \orgaddress{\street{Evanston, IL}, \city{City}, \postcode{60208}, \state{State}, \country{USA}}}

\affil[3]{\orgdiv{Electrical and Computer Engineering}, \orgname{George Mason University}, \orgaddress{\street{4400 University Dr, Fairfax}, \city{Virginia}, \postcode{22030}, \state{Virginia}, \country{USA}}}


\abstract{
%Neuromorphic computing requires compact and energy-efficient hardware neurons capable of reproducing the diverse firing behaviors observed in biological neural systems. 
Biological neurons exhibit diverse firing dynamics that enable adaptive and stimulus-dependent signalling, yet reproducing these dynamics in hardware has remained an enduring challenge. In this work, we present an Izhikevich-inspired reconfigurable neuron that co-designs voltage-controlled magnetic tunnel junction (V-MTJ) dynamics with CMOS circuitry. The proposed architecture combines V-MTJ excitability dynamics, enabled by a tunable energy landscape, with CMOS recovery dynamics to generate five distinct neuronal firing patterns with different spiking, bursting and response characteristics. Our results, based on measured V-MTJ characteristics and circuit simulations using commercial GlobalFoundries 22-nm FD-SOI CMOS technology, show an average energy consumption of 145.44 fJ per spike. Algorithmic simulations further show that these firing dynamics reduce inference spike activity by up to 88.6\% while maintaining baseline classification accuracy. These results highlight the potential of V-MTJ/CMOS reconfigurable neurons to reduce computational activity and enable compact, energy-efficient brain-inspired computing systems.}

\keywords{Voltage-controlled magnetic anisotropy,
Magnetic tunnel junction, Biologically plausible, Izhikevich neuron model, Neuromorphic hardware}



\maketitle

%%\section{Introduction}\label{sec1}

The human brain performs complex cognitive functions while consuming only about 20 W of power \cite{fang2023spikingjelly,gebregiorgis2025spike}. This efficiency arises from billions of interconnected neurons that communicate through precisely timed electrical spikes to encode and process information \cite{garcia2025spiking,eshraghian2023training,kudithipudi2025neuromorphic}. Beyond individual spikes, neurons exhibit diverse firing behaviors, including tonic and phasic spiking, adaptation, and bursting \cite{izhikevich2003simple,fortuna2023spiking,ceballos2025interleaved,alkabaa2022investigation}. Experimental electrophysiological recordings confirm this diversity through non-adapting and adapting spiking, delayed firing, and transient and persistent bursting \cite{komendantov2019quantitative}. Regular-spiking neurons use spike-frequency adaptation to encode changes in sustained inputs, while fast-spiking neurons support rapid and temporally precise information processing \cite{zare2021area,ganguly2024spike,nicola2024impact}. Bursting and phasic neurons further support robust information encoding and stimulus-onset detection, respectively \cite{xie2024neuronal,lee2023spike}.

The diverse firing dynamics of biological neurons have inspired spiking neural networks (SNNs), which use spike-based, event-driven processing for temporally precise and energy-efficient computation \cite{roy2019towards,campbell2022considerations,sutton2021spiking,yao2024spike,abaleke2026hardware,gou2025dynamic,fan2025multisynaptic}. However, reproducing rich neuronal dynamics in compact and energy-efficient hardware remains challenging. Biophysically detailed models such as Hodgkin–Huxley capture rich dynamics but require complex circuitry \cite{demirkol2011low,balubaid2022central}, whereas simpler leaky integrate-and-fire (LIF) models enable efficient implementations but cannot reproduce dynamics such as bursting and adaptation \cite{uludaug2024bio}. The Izhikevich model offers a balance between dynamical richness and computational efficiency, making it attractive for neuromorphic hardware \cite{andabili2025chaotic}.

Emerging nanoscale devices, including memristive, phase-change, ferroelectric, and spintronic technologies, exploit intrinsic nonlinear dynamics to implement neuronal and synaptic functionalities \cite{dutta2019biologically,tuma2016stochastic,yang2022spintronic}. Spintronic devices are particularly attractive because their magnetic dynamics provide nonlinearity and stochasticity, together with non-volatility, fast operation, and CMOS compatibility \cite{grollier2020neuromorphic,marrows2024neuromorphic,zhou2021prospect}. However, spintronic neurons have largely relied on spin-transfer torque (STT) switching, which requires high current densities and incurs substantial power overhead \cite{liang2020stochastic,tan2023backhopping,mehonic2024roadmap,vatajelu2017fully}. The current-driven operation of STT-MTJs also constrains voltage scaling, particularly for high-resistance junctions under scaled CMOS supply voltages \cite{liu2023binarized}, while high switching currents can induce self-heating and degrade device endurance and reliability \cite{mehonic2024roadmap}.

These limitations necessitate large current-driving access transistors and peripheral circuitry, offsetting some of the intrinsic advantages of spintronic devices \cite{liang2020stochastic}. Recent efforts have therefore explored voltage-controlled magnetic anisotropy (VCMA), which enables low-power control of voltage-controlled magnetic tunnel junctions (V-MTJs) without large drive currents. The voltage-dependent magnetic energy landscape of the V-MTJ enables tuning of the thermally activated P--AP dwell time, providing excitability dynamics that can be coupled with slower CMOS recovery dynamics to realize diverse neuronal firing behaviors. V-MTJs are also compatible with CMOS back-end-of-line (BEOL) integration \cite{duffee2026codesigned}, facilitating V-MTJ/CMOS co-integration. This device--circuit interaction provides a route toward run-time reconfigurable artificial neurons with diverse firing dynamics. Beyond biological fidelity, such reconfigurability can improve the computational efficiency of spike-based systems by controlling neuronal activity and spike-mediated communication \cite{moshruba2025izhikevich}.

\begin{figure}[!t]
    \centering
    \includegraphics[
        width=\textwidth,
        height=0.75\textheight,
        keepaspectratio
    ]{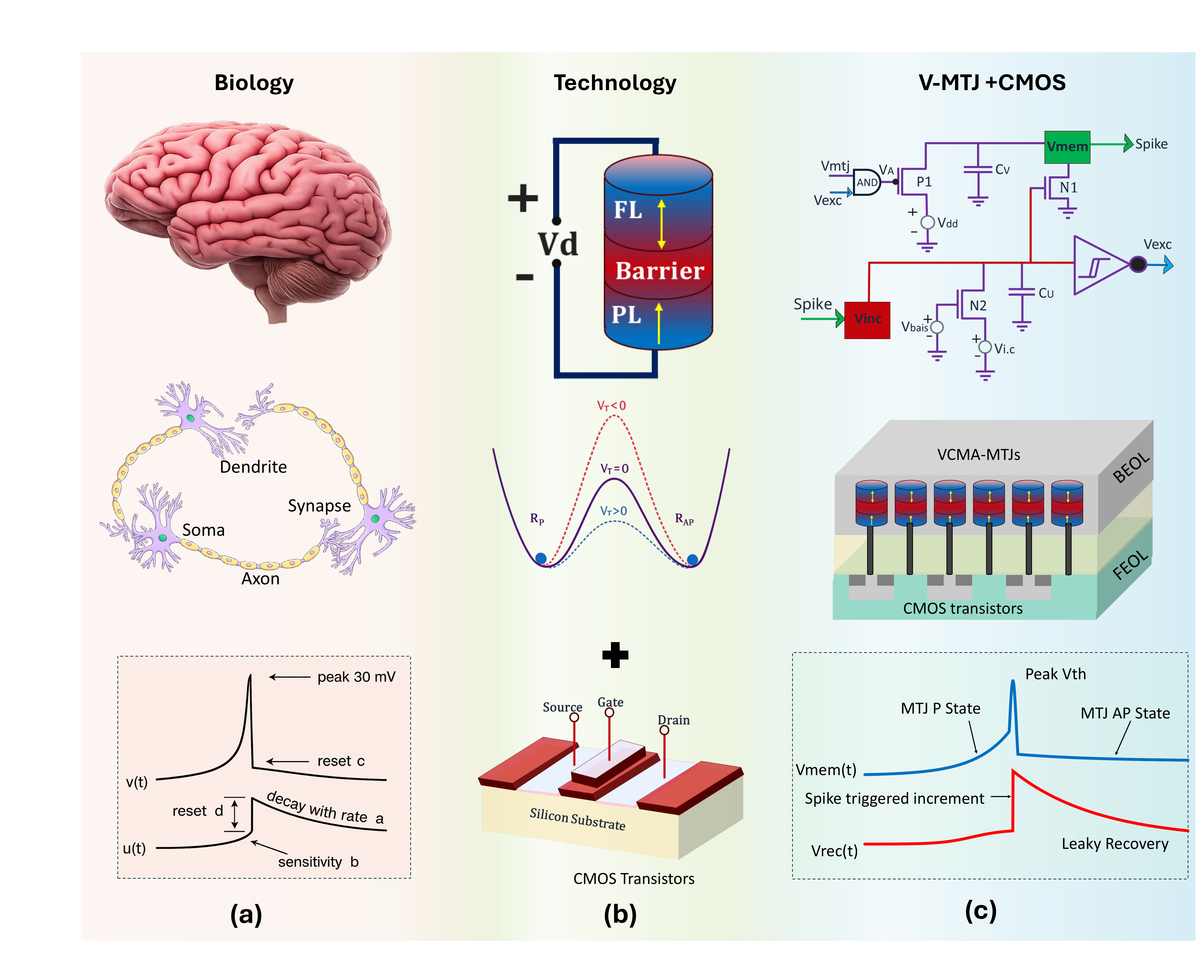}
    \caption{
    \textbf{V-MTJ/CMOS neuron architectural blocks:}
    (a) Biological brain.
    (b) 3D-integrated V-MTJ/CMOS neuron technology.
    (c) Circuit implementation.
    }
    \label{fig:introductory_image}
\end{figure}

The key contributions of this work are as follows: to the best of our knowledge, for the first time we propose a V-MTJ/CMOS co-designed spiking neuron inspired by the interaction of fast excitability and slow recovery variables in the Izhikevich neuron model (Fig. \ref{fig:introductory_image} a--c). Specifically, by combining V-MTJ dynamics that emulate fast excitability with CMOS adaptation circuitry that reproduces slow adaptation dynamics, the proposed neuron enables reconfigurable multi-pattern firing. Furthermore, the design realizes these firing behaviors within a single compact hardware framework without requiring separate circuit implementations or physical device resizing. We validate the proposed neuron using experimentally measured V-MTJ characteristics together with CMOS circuit simulations in commercial GlobalFoundries 22-nm FD-SOI technology. Finally, through software-level spiking neural network simulations, we further show that the rich dynamics realized by the proposed hardware can provide substantial temporal sparsity benefits: across four benchmark datasets, these firing modes achieve classification accuracy on par with (and in some cases exceeding) conventional LIF neurons while using up to 88.6\% fewer spikes.

%%\section{Results}\label{sec2}
%%In this section, we present the operating principle of the proposed VCMA-MTJ/CMOS neuron and discuss the resistance characteristics and  dwell time characteristics of the MTJ device. We then demonstrate a reconfigurable neuron circuit capable of reproducing various cortical spiking and bursting dynamics, including tonic spiking, phasic spiking, tonic bursting, phasic bursting, and spike-latency behavior.

\subsection*{V-MTJ switching as a voltage-gated neural excitability substrate}

\begin{figure}[!t]
    \centering
    \includegraphics[
        width=\textwidth,
        height=0.75\textheight,
        keepaspectratio
    ]{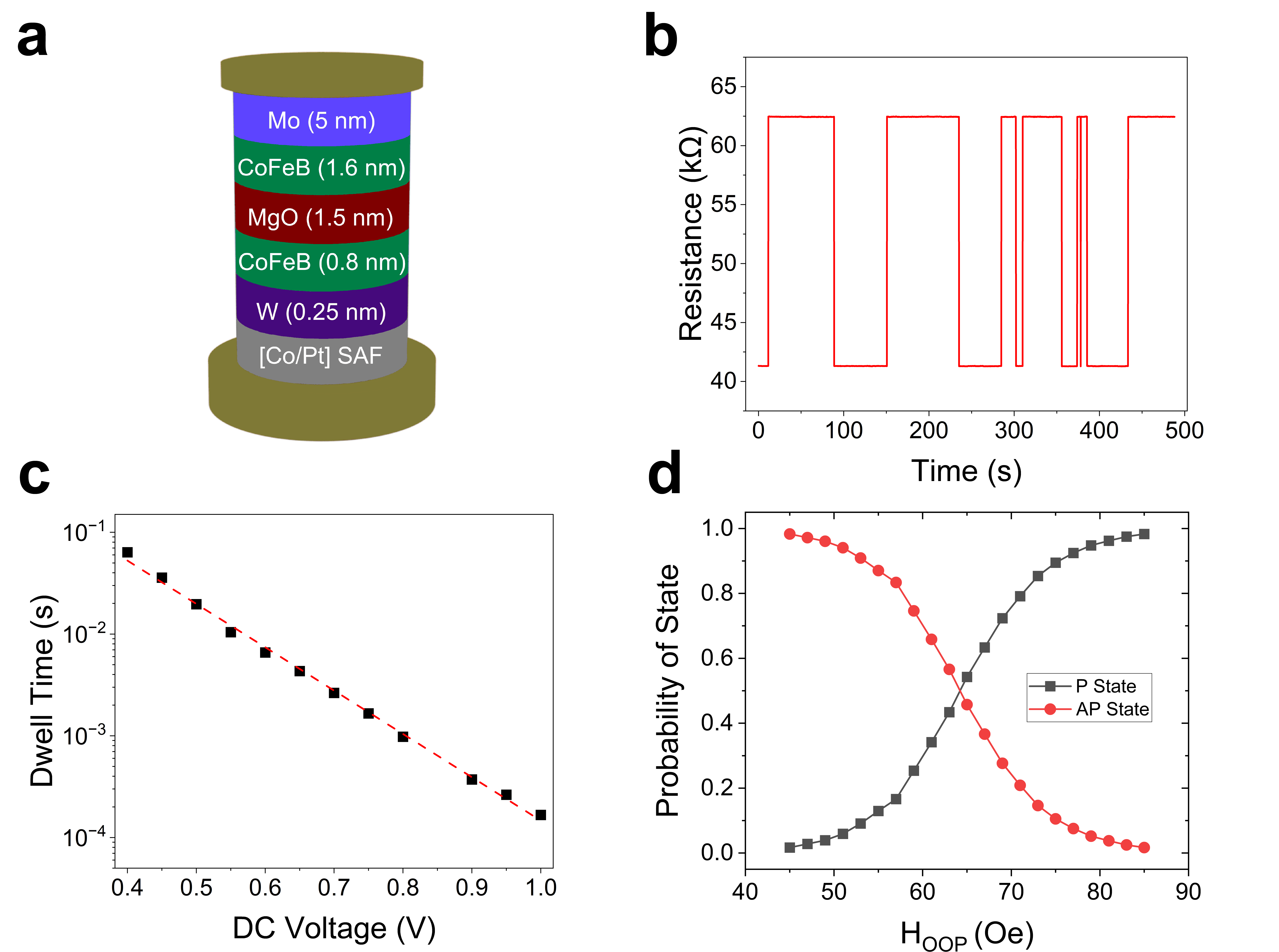}
    \caption{
    \textbf{V-MTJ characterization:}
    (a) Device stack.
    (b) Time trace of device resistance fluctuating due to thermal stochasticity.
    (c) Magnetization dwell time as function of voltage.
    (d) Probability of finding the device in the P or AP state at a given time as a function of external magnetic-field bias.
    }
    \label{fig:vcma-mtj_image}
\end{figure}

Biological neurons process information through changes in membrane potential. Excitatory synaptic input drives the membrane potential from its resting state toward a firing threshold, producing a spike followed by repolarization and recovery. The interaction between fast membrane-potential dynamics and slower recovery dynamics gives rise to a wide range of cortical firing behaviors, as captured by the four-parameter Izhikevich neuron model \cite{izhikevich2003simple}.

\begin{equation}
\frac{dv}{dt}=0.04v^{2}+5v+140-u+I
\label{eq:izhikevich1}
\end{equation}

\begin{equation}
\frac{du}{dt}=a(bv-u)
\label{eq:izhikevich2}
\end{equation}
When the membrane voltage reaches the spike-detection threshold
\begin{equation}
V \geq V_{\mathrm{threshold}},
\qquad
\begin{cases}
v \leftarrow c,\\
u \leftarrow u+d.
\end{cases}
\label{eq:izhikevich3}
\end{equation}

\noindent where $v$ is the membrane potential (mV), $u$ is the membrane recovery variable, $I$ is the input current, $a$ is the recovery time constant, \(V_{\mathrm{threshold}}\) is the  spike-detection threshold, $b$ is the sensitivity of the recovery variable $u$ to the membrane potential, $c$ is the membrane reset potential after a spike, and $d$ is the recovery increment after a spike. 

In the proposed neuron, the fast excitability mechanism is provided by an experimentally characterized V-MTJ. The perpendicular magnetic tunnel junction structure used in this work is shown in Fig.~\ref{fig:vcma-mtj_image}a. The device consists of a CoFeB/MgO/CoFeB stack. Application of a voltage across the MgO tunnel barrier modifies the interfacial magnetic anisotropy energy, thereby changing the effective magnetic energy barrier separating the parallel (P) and antiparallel (AP) states \cite{athas2025statistical}. Unlike conventional spin-transfer torque devices that require large switching currents, VCMA enables electric-field-driven control of the magnetic state, allowing the V-MTJ mean dwell time to be continuously tuned through the applied voltage \cite{amiri2012voltage}.

The resulting stochastic switching behavior is illustrated experimentally (Fig.~\ref{fig:vcma-mtj_image}b), which shows a representative resistance-time trace measured under a consonant applied bias  \(V_{\mathrm{T}}\). Thermal activation causes the free layer magnetization to randomly transition between the low-resistance P state and high-resistance AP state, producing random telegraph noise. The duration that the device remains in either magnetic state is referred to as the dwell time, which represents the characteristic lifetime of the magnetization before a thermally activated switching event occurs. The measured dwell time decreases approximately exponentially with increasing applied voltage (Fig.~\ref{fig:vcma-mtj_image}c), reflecting the progressive reduction of the magnetic energy barrier through the VCMA effect \cite{shao2022sub}. Consequently, higher applied voltages increase the switching rate and produce more frequent P/AP transitions. In addition to modifying the switching rate,  the external magnetic-field bias  also controls the relative occupancy of the two magnetic states (Fig.~\ref{fig:vcma-mtj_image}d). Near the balance operating condition, both states exhibit comparable occupation probabilities, whereas increasing the bias toward one magnetic state increases its residence probability while reducing that of the opposite state.

The proposed V-MTJ/CMOS neuron (Fig.~\ref{fig:introductory_image}b,c) operates through the interaction of V-MTJ excitability and CMOS recovery dynamics. In the antiparallel (AP) state, the high MTJ resistance \(R_{AP}\) suppresses the P1 charging pathway, keeping the membrane near its resting state. An applied voltage \(V_{\mathrm{T}}\) modifies the interfacial magnetic anisotropy through the VCMA effect, tuning the energy barrier and mean dwell time of the V-MTJ. When the MTJ is in the low-resistance parallel (P) state, the resulting excitation signal enables P1, allowing the membrane capacitor to charge toward the firing threshold, analogous to membrane depolarization, and generate a spike. During repeated spiking, each spike increases the recovery voltage, strengthening the recovery-controlled discharge and suppressing membrane excitation. This produces a silent or refractory interval analogous to the recovery dynamics and membrane hyperpolarization associated with reduced neuronal excitability in biological neurons.

\subsection*{Device-Circuit co-design for adaptive spiking and burst generation}
\begin{figure}[htbp]
    \centering
    \includegraphics[
        width=1.0\textwidth,
        keepaspectratio
    ]{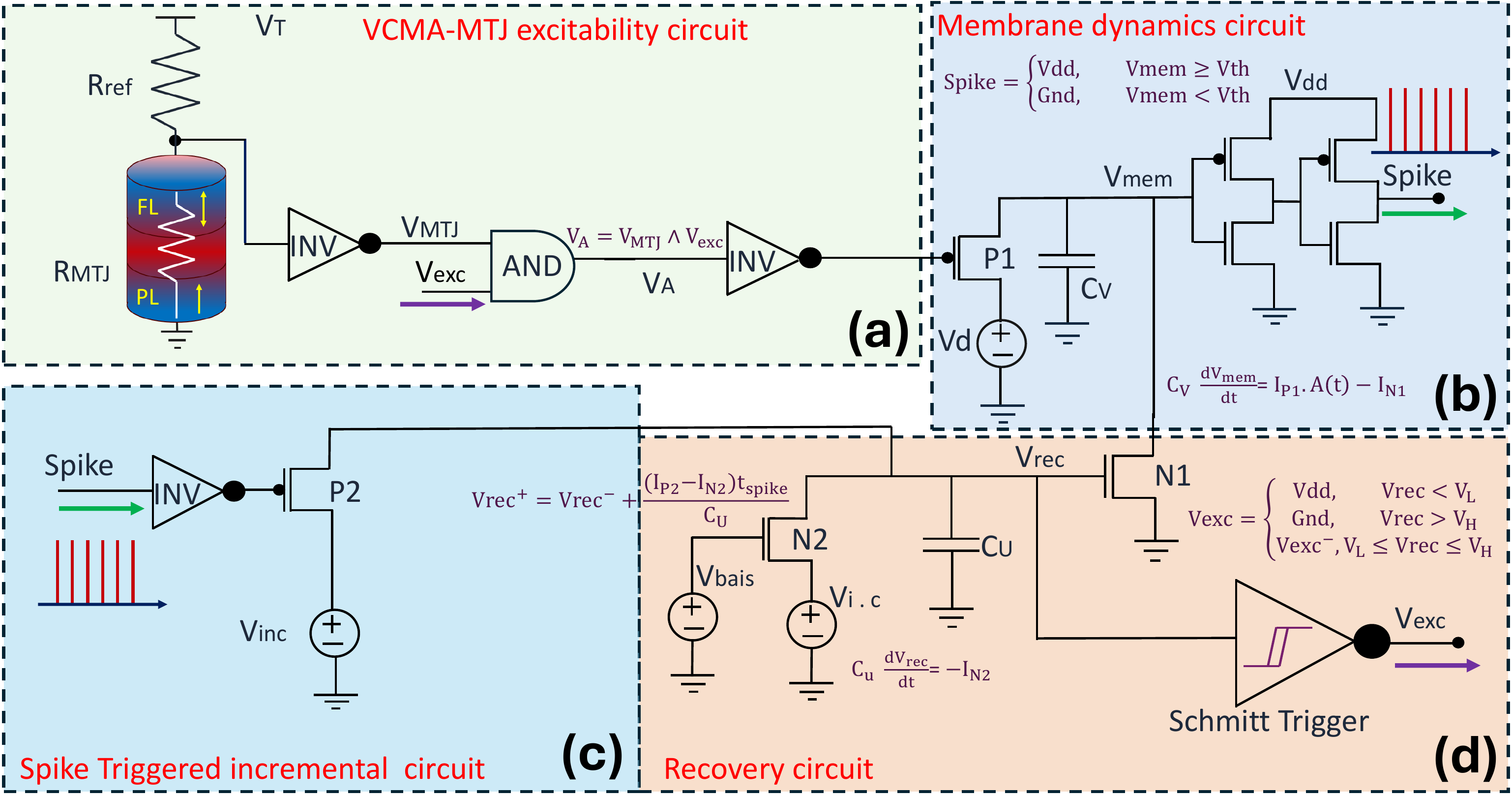}
    \caption{
\textbf{V-MTJ/CMOS neuron architectural blocks:}
(a) V-MTJ excitability circuit for generating the excitation window through voltage-controlled MTJ switching dynamics.
(b) Membrane dynamics circuit for membrane charging, threshold integration, and spike generation.
(c) Spike-triggered increment circuit for recovery accumulation and firing adaptation.
(d) Recovery circuit for implementing recovery-controlled leakage and temporal adaptation dynamics.
}
    \label{fig:complete_circuit}
\end{figure}

Here, we discuss the complete V-MTJ/CMOS neuron architecture as shown in Fig.~\ref{fig:complete_circuit}, including the interaction between the V-MTJ excitability circuit, membrane dynamics circuit, recovery circuit, and spike triggered increment circuit. We further introduce the mathematical framework governing the interaction between the fast excitability mechanism and the slow recovery dynamics responsible for adaptation and temporal firing regulation in the proposed neuron.

The proposed V-MTJ/CMOS neuron operates through the interaction between fast excitation dynamics generated by the V-MTJ excitability circuit (Fig.~\ref{fig:complete_circuit}a) and slow recovery dynamics implemented by the CMOS adaptation circuit (Fig.~\ref{fig:complete_circuit}d). The V-MTJ output \(V_{\mathrm{MTJ}}\) is combined with the recovery-controlled enable signal \(V_{\mathrm{exc}}\) generated by the Schmitt-trigger-based recovery circuit. The resulting AND-gate output, denoted by \(V_{\mathrm{A}}\), controls the charging activity of the membrane dynamics circuit (Fig.~\ref{fig:complete_circuit}b) and is expressed as ~\ref{eqn:1}:

\begin{equation}
V_{\mathrm{A}} = V_{\mathrm{MTJ}} \land V_{\mathrm{exc}}
\label{eqn:1}
\end{equation}

For clarity, the physical node $V_A$ is represented by the dimensionless variable $A(t)$, where $A(t)=1$ when \(V_{\mathrm{A}}\) =  \(V_{\mathrm{DD}}\) (0.8V) and $A(t)=0$ when \(V_{\mathrm{A}}\) =  GND. Throughout this work, $A(t)$ denotes the excitation state of the neuron, while $V_A$ refers to the corresponding physical circuit node. The output voltage $V_{\mathrm{MTJ}}$ is near  $\mathrm{GND}$ when the MTJ is in the high-resistance state  and near  \(V_{\mathrm{DD}}\) when it is in the low-resistance state.

The \(V_{\mathrm{MTJ}}\) output depends on the intrinsic mean dwell time of the V-MTJ, which can be tuned by the applied voltage through the VCMA effect. When the MTJ is in the low-resistance state, the voltage-divider output is inverted to produce a high \(V_{\mathrm{MTJ}}\) signal. With the recovery-enable signal \(V_{\mathrm{exc}}\) also high, the AND gate drives \(V_A\) to \(V_{DD}\). The \(V_A\) signal is subsequently passed through an inverter and applied to the gate of the PMOS transistor P1. The inverted low output turns P1 on, enabling current to flow from the supply voltage \(V_d\) into the membrane capacitor \(C_v\). Conversely, when either \(V_{\mathrm{MTJ}}\) or \(V_{\mathrm{exc}}\) is low, \(V_A\) is driven to \(\mathrm{GND}\), resulting in a high voltage at the gate of P1. This turns P1 off and suppresses membrane charging.

Accordingly, \(A(t)=1\) represents an active excitation state in which P1 contributes to membrane charging, whereas \(A(t)=0\) suppresses the P1 charging path. The membrane voltage \(V_{\mathrm{mem}}\) is determined by the interaction between the excitation-driven charging current through P1 and the recovery-controlled discharge current through N1, whose gate is driven by \(V_{\mathrm{rec}}\). The membrane-node dynamics are described by the nodal current balance in Equation~\ref{eq:4}:

\begin{equation}
C_v \frac{dV_{\mathrm{mem}}}{dt}
=
I_{P1} A(t) - I_{N1}
\label{eq:4}
\end{equation}

where \(C_v\) is the membrane capacitor, \(I_{P1}\) is the charging current from the supply voltage through transistor P1, \(A(t)\) represents the effective excitation state of the neuron, and \(I_{N1}\) is the recovery-controlled discharge current through transistor N1, whose gate is driven by the recovery voltage \(V_{\mathrm{rec}}\).

Equation ~\ref{eq:4} captures the interaction between the fast excitation and slow recovery processes that characterize the Izhikevich neuron model as reported in \cite{izhikevich2003simple}. In the proposed hardware realization, the membrane capacitor charges when the excitation state A(t) is active while the recovery-controlled transistor N1 provides gradual negative feedback through \(V_{\mathrm{rec}}\), thereby regulating spike generation and adaptation.
During the inactive state of the MTJ (\(V_{\mathrm{MTJ}}\) = 0), the effective excitation signal A(t) remains disabled, keeping the membrane voltage below the firing threshold and corresponding to complete neuronal silence.

During the active state of the MTJ (\(V_{\mathrm{MTJ}}\) = 1), the effective excitation signal A(t) remains active whenever the recovery-controlled gating signal \(V_{\mathrm{exc}}\) is enabled, allowing the membrane excitation to be regulated by the recovery variable \(V_{\mathrm{rec}}\). Initially, the recovery voltage remains low, resulting in a weak pull-down action through transistor N1 and allowing the membrane voltage to increase toward the firing threshold voltage and generate spike signal.
The generated spike drives the spike-triggered incremental circuit (Fig.~\ref{fig:complete_circuit}c), producing an increment in the recovery voltage \(V_{\mathrm{rec}}\). Each generated spike injects additional charge into the recovery capacitor $C_u$ through the spike-triggered incremental circuit, increasing the recovery voltage $V_{\mathrm{rec}}$ during the spike. The spike-triggered recovery increment can be expressed as:
 
\begin{equation}
V_{\mathrm{rec}}^{+}
=
V_{\mathrm{rec}}^{-}
+
\frac{\left(I_{P2}-I_{N2}\right)t_{\mathrm{spike}}}{C_u}
\label{eqn:recovery_update}
\end{equation}

where $V_{\mathrm{rec}}^{-}$ and $V_{\mathrm{rec}}^{+}$ represent the recovery voltage before and after a spike event, respectively. $I_{P2}$ is the spike-triggered charging current through transistor P2, $I_{N2}$ is the discharge current through transistor N2, $t_{\mathrm{spike}}$ is the spike duration, and $C_u$ is the recovery capacitance. This equation describes the spike-triggered update of the recovery voltage through the combined action of P2 and N2. During the spike-triggered increment, P2 charges \(C_u\), increasing \(V_{\mathrm{rec}}\), while N2 provides a simultaneous discharge path. After the spike, P2 turns off and N2 provides the slow recovery discharge, functionally representing the recovery dynamics of the Izhikevich model \cite{izhikevich2003simple}, which may be expressed as:
 
\begin{equation}
C_u\frac{dV_{\mathrm{rec}}}{dt}=-I_{N2}
\label{eqn:recovery_decay}
\end{equation}

where $I_{N2}$ is the discharge current through transistor N2, which establishes the slow recovery dynamics by discharging the recovery capacitor between spike events. During repeated spike activity, each spike increases $V_{\mathrm{rec}}$ through the spike-triggered incremental circuit. The increased $V_{\mathrm{rec}}$ strengthens the pull-down action of transistor N1, reducing $V_{\mathrm{mem}}$ and suppressing membrane excitation. Between spike events, N2 discharges the recovery capacitor, allowing $V_{\mathrm{rec}}$ to decrease.

The membrane voltage $V_{\mathrm{mem}}$ is sensed by a two-inverter thresholding stage that serves as the spike-generation circuit. When $V_{\mathrm{mem}}$ reaches or exceeds the threshold voltage $V_{\mathrm{th}}$, the circuit generates a spike according to:

\begin{equation}
\mathrm{Spike}(t)=
\begin{cases}
V_{\mathrm{dd}}, & V_{\mathrm{mem}} \geq V_{\mathrm{th}} \\
Gnd, & V_{\mathrm{mem}} < V_{\mathrm{th}}
\end{cases}
\label{eq:3}
\end{equation}

This threshold-driven spike generation mechanism directly maps onto the depolarization-to-firing transition observed in biological neurons, where sufficient membrane excitation initiates rapid spike activity. The membrane capacitor integrates excitation over time, allowing the temporal evolution of the membrane voltage to regulate spike timing, firing dynamics, and neuronal response behavior. In the proposed architecture, these dynamics can be tuned through the membrane charging current controlled by the supply voltage and the recovery dynamics regulated by $V_{\mathrm{bias}}$ and $V_{\mathrm{I.C}}$ through transistor N2. The gate of N2 is biased by $V_{\mathrm{bias}}$, while its source is driven by the externally applied $V_{\mathrm{I.C}}$ pulse to modulate the recovery dynamics and enable different temporal firing responses. The recovery capacitor $C_u$, together with the discharge path through N2, determines the recovery timescale and thereby influences the temporal adaptation of the neuron.

The recovery voltage \(V_{\mathrm{rec}}\) is also connected to the input of the Schmitt trigger, which generates the recovery-controlled enable signal  \(V_{\mathrm{exc}}\). The hysteretic switching thresholds of the Schmitt trigger regulate the transitions between active and inactive firing states. While the recovery voltage remains at or below the lower hysteretic switching threshold $V_L$, the Schmitt trigger maintains $V_{\mathrm{exc}}$ at \(V_{\mathrm{DD}}\), allowing the membrane capacitor to charge when the MTJ is in the low-resistance parallel (P) state. With successive spike activity, each spike increases the recovery voltage $V_{\mathrm{rec}}$ through spike-triggered charge injection. When $V_{\mathrm{rec}}$ exceeds the upper hysteretic switching threshold $V_H$, the Schmitt trigger drives $V_{\mathrm{exc}}$ to $\mathrm{GND}$, disabling the excitation pathway, stopping further membrane charging, and suppressing subsequent spike generation, thereby placing the neuron in a refractory state. Within the hysteresis region ($V_L \leq V_{\mathrm{rec}} \leq V_H$), $V_{\mathrm{exc}}$ retains its previous state until either switching threshold is crossed, thereby ensuring stable hysteretic control of the excitation signal. The Schmitt-trigger behavior may be approximated as:

\begin{equation}
 V_{\mathrm{exc}}=
\begin{cases}
V_{DD}, & V_{\mathrm{rec}}\leq V_L\\
Gnd, & V_{\mathrm{rec}}\geq V_H\\
\text{previous state}, & V_L<V_{\mathrm{rec}}<V_H
\end{cases}
\label{eq:7}
\end{equation}
where  $V_{L}$ and $V_{H}$ correspond to the lower and upper hysteretic switching thresholds, respectively. The hysteretic nature of the gating mechanism prevents unstable transitions near the recovery threshold and enables stable burst termination and recovery behavior.

Together, the coupled operation of the V-MTJ excitability circuit, membrane dynamics circuit, recovery circuit, and spike-triggered increment block establishes a compact hardware realization of fast–slow neuronal dynamics. The V-MTJ provides the fast excitation timescale responsible for neuronal activation and spike generation, while the CMOS recovery circuitry introduces the slower adaptation dynamics that regulate excitability over time. Because the recovery voltage changes dynamically through spike-triggered accumulation and gradual leakage, the neuron regulates its firing activity based on both its current excitation level and its prior spike history, rather than relying only on a fixed firing threshold. By tuning the MTJ excitation window, recovery leakage rate, and recovery incremental voltage, the proposed architecture enables reconfigurable temporal firing behaviors within a single compact hardware architecture without requiring separate circuit implementations or transistor resizing.

\subsection*{Firing patterns generated by the proposed V-MTJ/CMOS neuron model}
\begin{figure}[!htbp]
\centering

%---------------- First row ----------------%
\begin{minipage}{0.44\textwidth}
    \centering
    \includegraphics[
        width=\linewidth,
        height=0.289\textheight,
        keepaspectratio
    ]{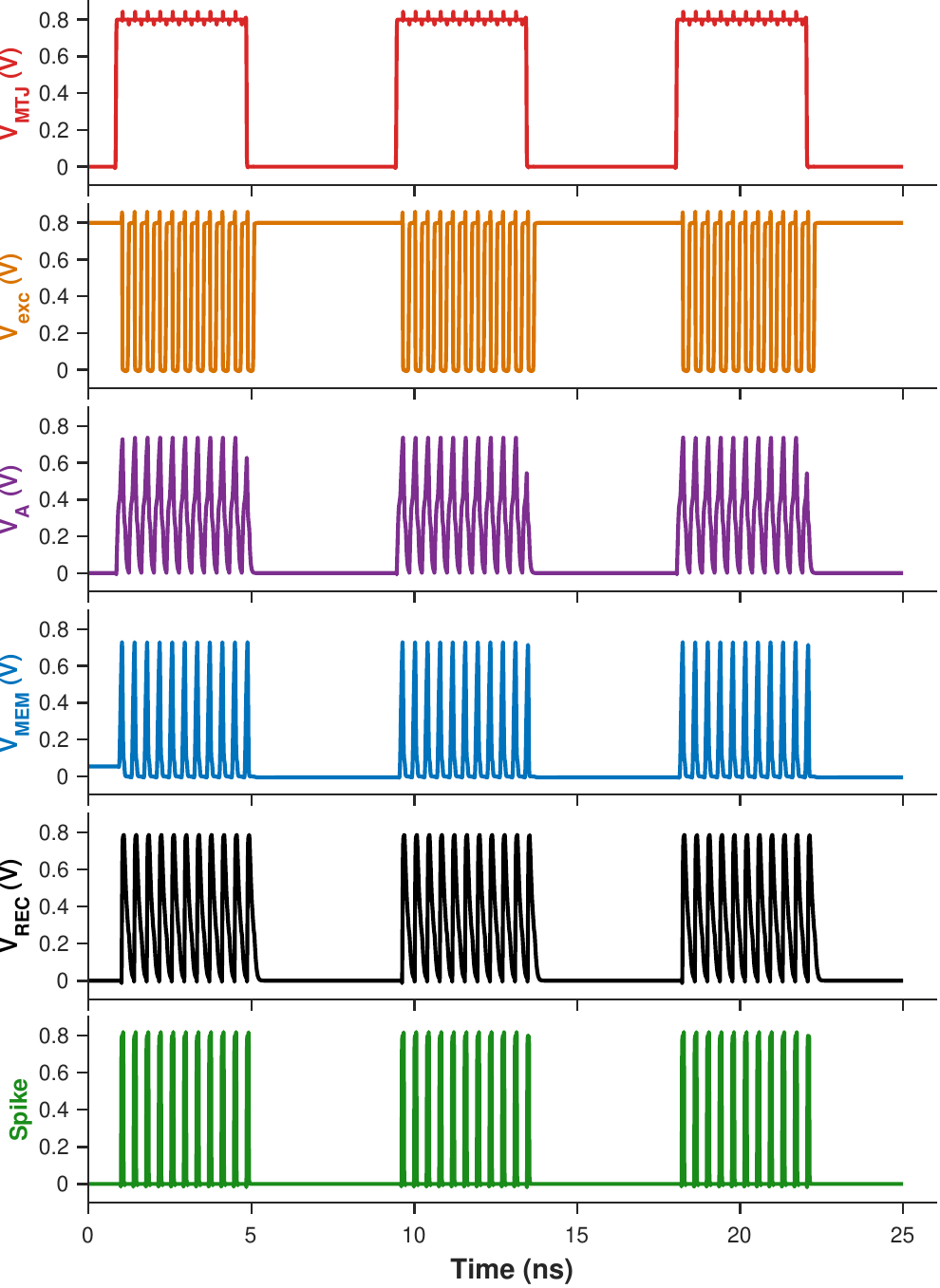}

    \scriptsize (a) Tonic bursting
\end{minipage}
\hfill
\begin{minipage}{0.44\textwidth}
    \centering
    \includegraphics[
        width=\linewidth,
        height=0.289\textheight,
        keepaspectratio
    ]{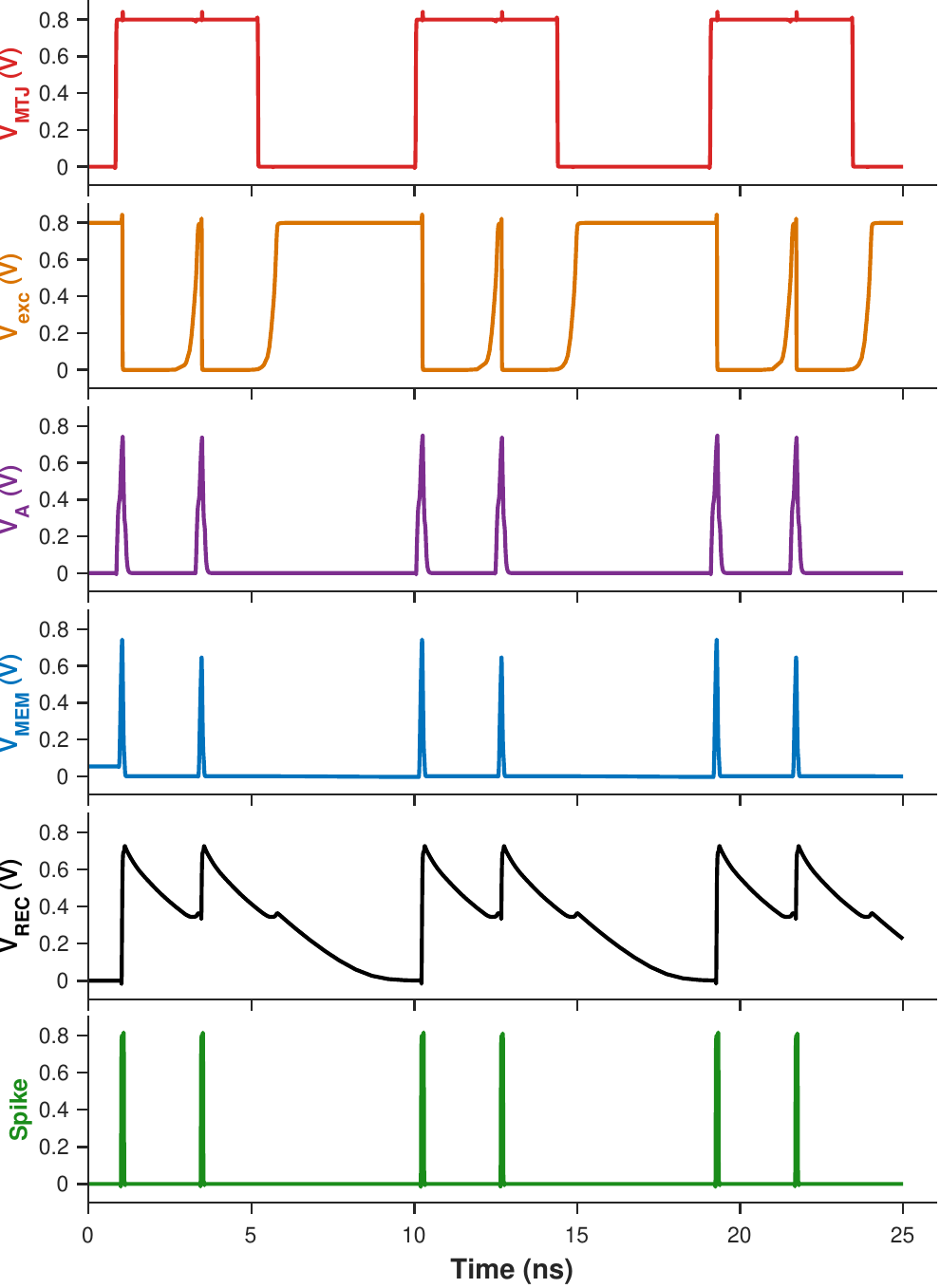}

    \scriptsize (b) Tonic spiking
\end{minipage}

\vspace{0.05cm}

%---------------- Second row ----------------%
\begin{minipage}{0.44\textwidth}
    \centering
    \includegraphics[
        width=\linewidth,
        height=0.289\textheight,
        keepaspectratio
    ]{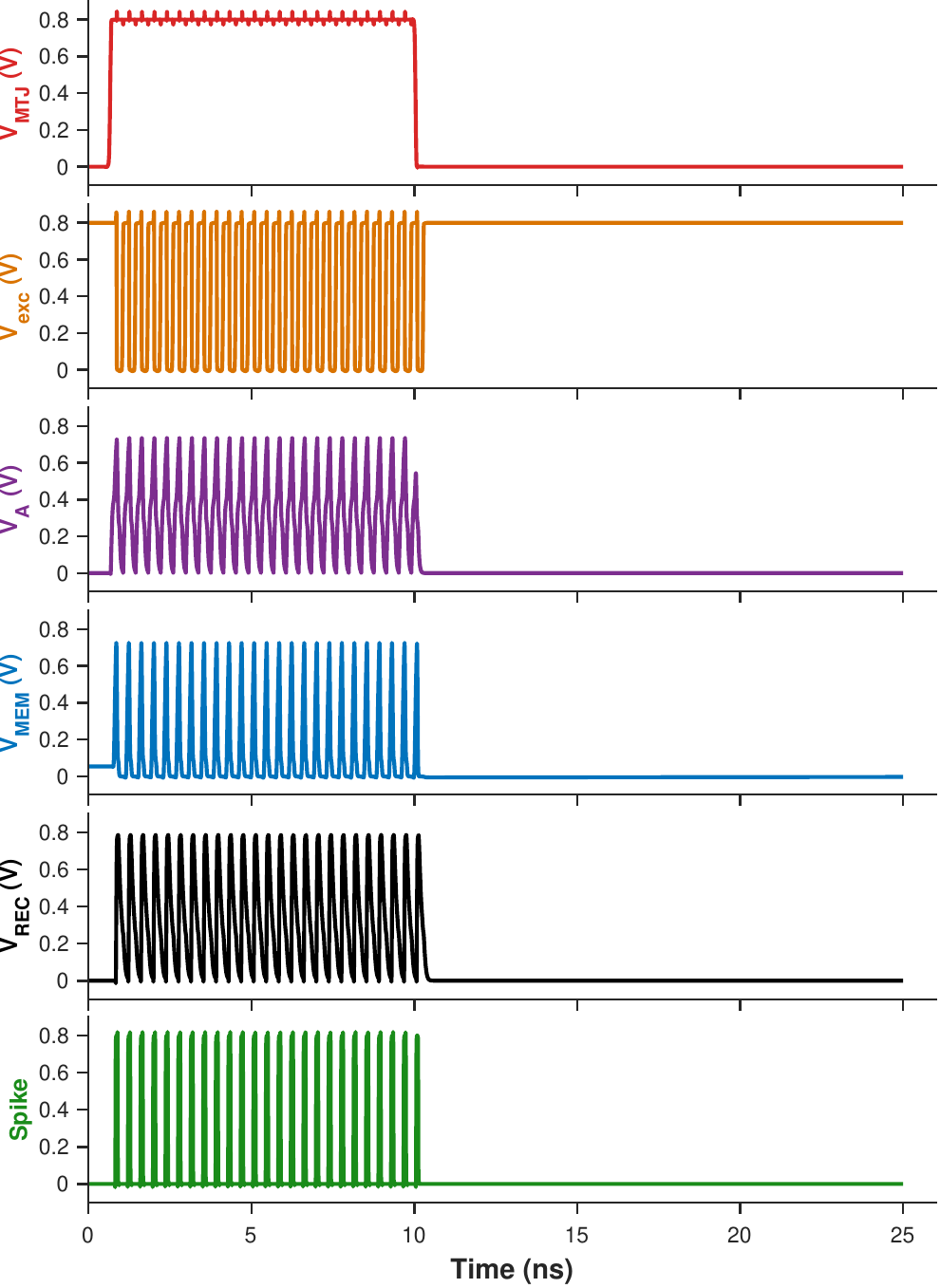}

    \scriptsize (c) Phasic bursting
\end{minipage}
\hfill
\begin{minipage}{0.44\textwidth}
    \centering
    \includegraphics[
        width=\linewidth,
        height=0.289\textheight,
        keepaspectratio
    ]{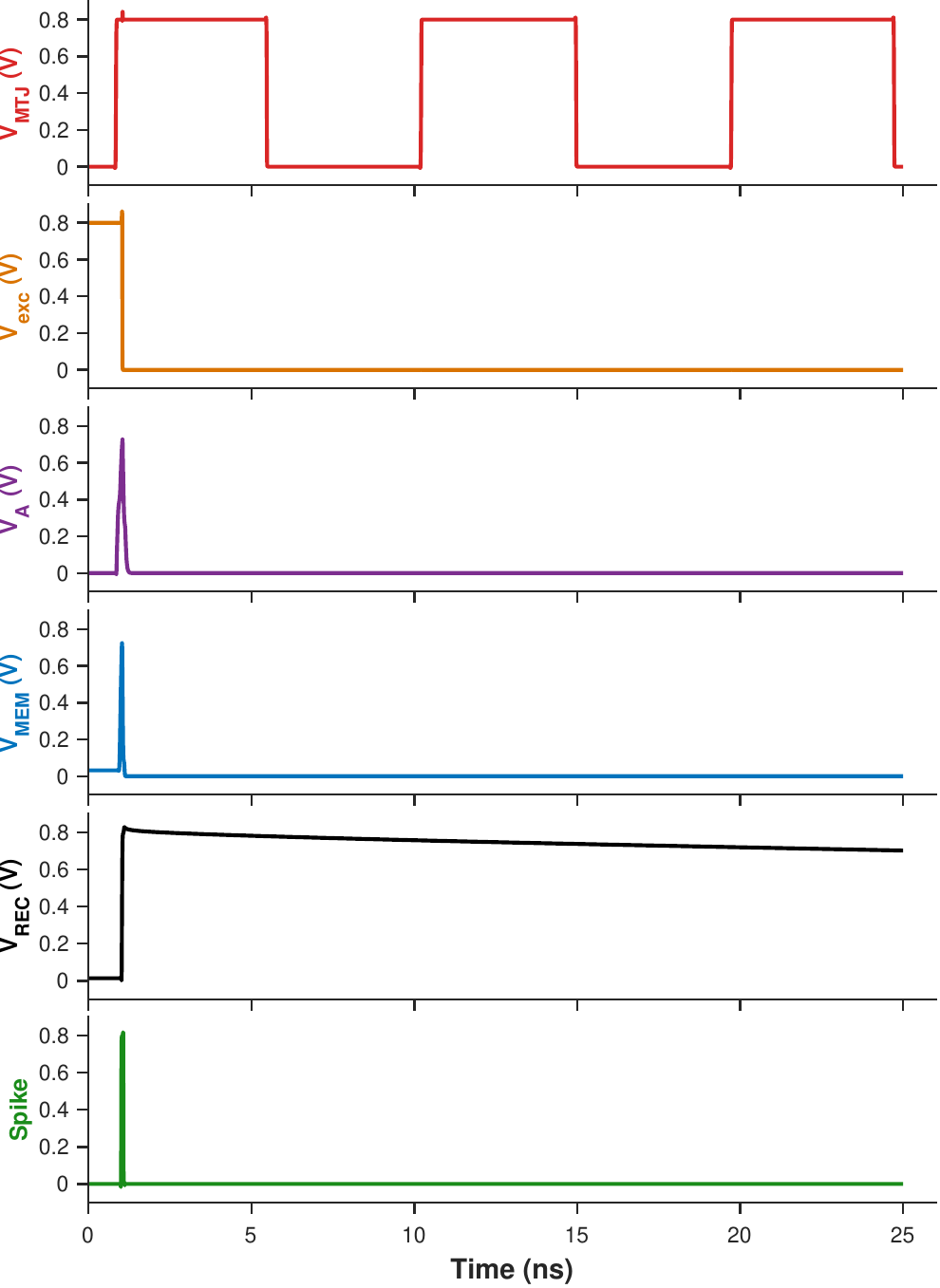}

    \scriptsize (d) Phasic spiking
\end{minipage}

\vspace{0.05cm}

%---------------- Third row ----------------%
\begin{minipage}{0.44\textwidth}
    \centering
    \includegraphics[
        width=\linewidth,
        height=0.289\textheight,
        keepaspectratio
    ]{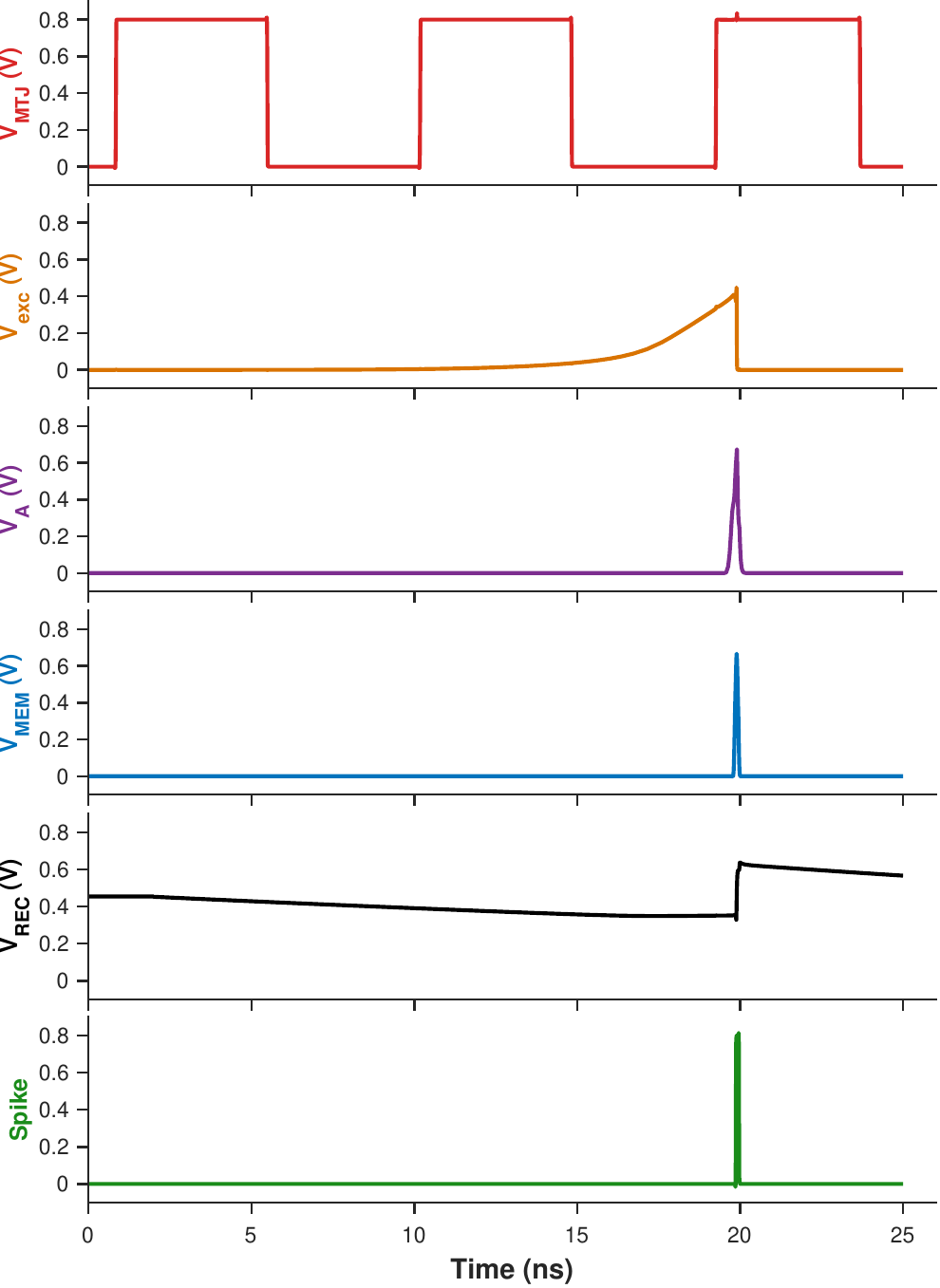}

    \scriptsize (e) Spike latency
\end{minipage}

\caption{\textbf{Simulated firing behaviors of the proposed V-MTJ/CMOS neuron, demonstrating its ability to emulate diverse biologically inspired firing patterns.} (a) Tonic bursting, characterized by repetitive burst events. (b) Tonic spiking, exhibiting sustained periodic spike generation. (c) Phasic bursting, producing a single burst event. (d) Phasic spiking, producing a single spike. (e) Spike latency, characterized by a delayed spike.}
\label{fig:firing_patterns}

\end{figure}

The proposed V-MTJ/CMOS neuron reproduces multiple cortical firing patterns within a single compact hardware architecture (Fig.~\ref{fig:firing_patterns}), where the observed firing behavior is determined through voltage-controlled tuning of the excitation and recovery dynamics. Specifically, the neuronal response is governed by the interaction between the recovery bias voltage $V_{\mathrm{bias}}$, the recovery pulse voltage $V_{\mathrm{I.C}}$, the V-MTJ control voltage $V_T$, and the spike-triggered increment voltage $V_{\mathrm{inc}}$. Together, these parameters regulate the recovery voltage $V_{\mathrm{rec}}$, membrane excitation, and recovery timescale, thereby enabling different neuronal firing behaviors. Consequently, each firing pattern emerges from a distinct operating region within the coupled excitation–recovery dynamics and maps directly onto the firing modes described in the Izhikevich neuron firing model. Note, due to the intrinsic speed of V-MTJs and associated CMOS circuits, the proposed hardware reproduces these biologically inspired behaviors at accelerated nanosecond timescales compared with millisecond-scale biological neurons.

The V-MTJ excitability circuit establishes the active spiking window of the neuron through voltage-driven switching between the antiparallel (AP) and parallel (P) MTJ states. During the active P-state interval, while the recovery-controlled enable signal \(V_{\mathrm{exc}}\) remains active, the membrane capacitor integrates excitation toward the firing threshold and enables spike generation. Conversely, during the AP-state interval, the excitation pathway is disabled, preventing membrane charging and suppressing spike generation, while the recovery variable gradually decays through the leaky recovery circuit. The V-MTJ/CMOS neuron was simulated using a V-MTJ model derived from experimental measurements and CMOS circuitry implemented in GlobalFoundries 22-nm FD-SOI technology. The membrane and recovery capacitances, $C_v$ and $C_u$, respectively, were implemented using metal--oxide--metal (MOM) capacitors, enabling compact on-chip realization of the membrane and recovery dynamics. The proposed neuron exhibits an average energy consumption of approximately 145.44 fJ per spike across the five demonstrated firing modes. Tonic bursting achieves the lowest average energy of 8.82 fJ/spike, whereas the spike-latency mode exhibits the highest energy consumption of 515.4 fJ/spike.

\subsubsection*{Tonic bursting}

Tonic bursting occurs when the recovery voltage increases during spike generation while also decaying sufficiently between successive burst events. During the MTJ P-state, the neuron enters the active excitation phase, where the low MTJ resistance enables membrane charging and repeated spike generation. Each generated spike injects a moderate amount of additional charge into the recovery capacitor \(C_u\) through the spike-triggered increment circuit, causing \(V_{\mathrm{rec}}\) to rise during the spike. The increased \(V_{\mathrm{rec}}\) strengthens the recovery-controlled discharge through N1, pulling down \(V_{\mathrm{mem}}\) and suppressing membrane excitation. As \(V_{\mathrm{rec}}\) subsequently decreases, the recovery-controlled gating restores the excitation pathway, allowing \(V_{\mathrm{mem}}\) to charge again and generate the next spike. These repeated excitation--recovery cycles generate the sequence of spikes while the MTJ remains in the P state. When the MTJ transitions to the AP state, the \(V_{\mathrm{MTJ}}\) excitation signal is removed, terminating the burst and initiating the silent phase.

A recovery bias voltage \(V_{\mathrm{bias}}\) within the range of 600 – 650 mV increases the drive strength of transistor N2, enabling rapid discharge of \(V_{\mathrm{rec}}\) between successive spikes during the MTJ P-state. During the MTJ AP state, membrane charging is terminated while \(V_{\mathrm{rec}}\) decays through transistor N2. Once \(V_{\mathrm{rec}}\) falls below the lower Schmitt-trigger threshold, \(V_{\mathrm{exc}}\) becomes active again, enabling membrane charging when the MTJ subsequently enters the P state. The applied voltage \(V_T\) tunes the mean dwell time of the MTJ through the VCMA effect, thereby influencing the duration of the MTJ state intervals. Consequently, the interaction between the MTJ state dynamics and the excitation--recovery mechanism produces repeated spike bursts separated by silent intervals, characteristic of tonic bursting behaviour (Fig. 4a).

\subsubsection*{Tonic spiking}

Tonic spiking occurs when moderate $V_{\mathrm{inc}}$ and $V_{\mathrm{bias}}$ levels establish a regulated interaction between spike-triggered recovery increments and recovery discharge during the active MTJ excitation window. During the MTJ P-state, the low MTJ resistance enables membrane charging, allowing the membrane voltage $V_{\mathrm{mem}}$ to reach the firing threshold. Each generated spike injects additional charge into the recovery capacitor $C_u$ through the spike-triggered incremental circuit, increasing the recovery voltage $V_{\mathrm{rec}}$ during the spike. Between spikes, $V_{\mathrm{rec}}$ decreases through the N2 discharge path. With a moderate spike-triggered increment, the resulting excitation--recovery interaction supports repeated membrane charging and spike generation, producing tonic spiking.

A recovery bias voltage in the range of 300 - 350 mV allows the transistor \(N2\) to slowly discharge the recovery voltage \(V_{\mathrm{rec}}\), preventing excessive recovery accumulation during spike generation. As a result, transistor \(N1\) does not strongly discharge the membrane node, allowing repeated spike generation throughout the active MTJ P-state interval. Following transition into the AP-state, the high MTJ resistance inhibits membrane charging while \(V_{\mathrm{rec}}\) gradually decays through transistor \(N2\) before the next excitation cycle begins. The neuron therefore exhibits sustained tonic spiking behavior (Fig. 4b).

\subsubsection*{Phasic Bursting}

For the phasic bursting mode, the measured V-MTJ characteristics exhibit a higher state occupancy probability and longer dwell time in the low-resistance parallel (P) state than in the high-resistance antiparallel (AP) state. The applied voltage \(V_T\) tunes the mean dwell time of the MTJ through the VCMA effect. As a result, the MTJ initially remains in the excitatory P state long enough to sustain membrane charging and generate a burst of spikes. During the burst, each spike injects charge into the recovery capacitor \(C_u\), causing \(V_{\mathrm{rec}}\) to rise during the spike. The increased \(V_{\mathrm{rec}}\) strengthens the recovery-controlled discharge through N1, pulling down \(V_{\mathrm{mem}}\) and suppressing membrane excitation. As \(V_{\mathrm{rec}}\) subsequently decreases, the excitation pathway is restored, allowing the next spike to occur while the MTJ remains in the P state. When the MTJ subsequently transitions to the AP state, the excitation pathway is disabled, membrane charging stops, and no further spikes are generated. Thus, the neuron produces a single phasic burst followed by a silent phase, characteristic of phasic bursting behaviour (Fig. 4c).

\subsubsection*{Phasic spiking}

Phasic spiking occurs when the spike-triggered increment voltage \(V_{\mathrm{inc}}\) is sufficiently high while the recovery bias voltage \(V_{\mathrm{bias}}\) remains low during the active MTJ excitation window. During the MTJ P-state, the low MTJ resistance enables membrane charging and allows the membrane voltage \(V_{\mathrm{mem}}\) to reach the firing threshold, generating a spike event. However, because the recovery increment voltage \(V_{\mathrm{inc}}\) is high, the generated spike injects a large amount of charge into the recovery capacitor \(C_u\), causing the recovery voltage \(V_{\mathrm{rec}}\) to increase rapidly during firing activity.

A low recovery bias voltage in the range of 150 – 200 mV reduces the pull-down strength of transistor  N2, resulting in a lower discharge rate of \(V_{\mathrm{rec}}\), thereby prolonging the retention of the recovery voltage \(V_{\mathrm{rec}}\) following the initial spike event. As \(V_{\mathrm{rec}}\) rises, transistor \(N1\) strongly discharges the membrane voltage, preventing further threshold crossings even though the MTJ remains in the active P-state. Because the recovery voltage decays slowly between successive MTJ P-AP transitions, the membrane node remains strongly suppressed by transistor \(N1\), preventing the membrane voltage from reaching the firing threshold during subsequent excitation cycles. The neuron therefore produces only a single spike response, characteristic of phasic spiking behavior (Fig. 4d).

\subsubsection*{Spike latency}

Spike-latency behavior occurs when the recovery voltage $V_{\mathrm{rec}}$ is initially charged by the externally applied recovery-control pulse $V_{\mathrm{I.C}}$. The elevated $V_{\mathrm{rec}}$ drives the Schmitt-trigger output $V_{\mathrm{exc}}$ low, forcing the AND-gate output $V_A$ low. The subsequent inverter produces a high gate voltage at P1, turning P1 off and disabling the membrane-charging pathway. At the same time, the elevated $V_{\mathrm{rec}}$ strengthens the N1 pull-down action, further suppressing the rise of $V_{\mathrm{mem}}$. Consequently, the membrane voltage cannot immediately reach the firing threshold.

With $V_{\mathrm{bias}}$ in the range of 150 – 200~mV, N2 provides a weak discharge path, allowing $V_{\mathrm{rec}}$ to decrease slowly. When $V_{\mathrm{rec}}$ falls below the lower Schmitt-trigger threshold $V_L$, $V_{\mathrm{exc}}$ switches high, re-enabling the excitation pathway. If the MTJ is in the P-state, the AND-gate output $V_A$ becomes high. The subsequent inverter produces a low gate voltage at P1, turning P1 on and allowing the membrane capacitor to charge. Once $V_{\mathrm{mem}}$ reaches the firing threshold, the neuron generates a delayed spike, producing the spike-latency behavior (Fig.~4e).

%---------------------------------------------------------------------------
% Software Section
%---------------------------------------------------------------------------

% NOTE: I HAVE ALSO INCLUDED A DOUBLE COLUMN VERSION OF THE FIGURE WHICH IS NARROWER JUST IN CASE!

% Software emulation of multi-pattern hardware dynamics.} Simulated neural responses of the adaptive exponential (AdEx) integrate-and-fire software proxy under a constant input stimulus, reproducing the (a) tonic bursting and (b) phasic bursting behaviors characteristic of the proposed hardware.

 \subsection*{Computational implications of multi-pattern firing} 

% Hamed: Please check if it's better to be a section, subsection, or subsubsection

% We assess the functional significance of rich neuronal dynamics beyond their hardware realization by simulating tonic-bursting and phasic-bursting responses using the adaptive exponential (AdEx) integrate-and-fire framework. This framework serves as our software proxy, accurately capturing the same fast-excitation and slow-recovery structure characteristic of the proposed device \cite{brette2005adaptive,naud2008firing}. As illustrated in Figure~\ref{fig:AdEx_Response}, the algorithmically modeled neurons successfully reproduce the distinct bursting behaviors generated by the hardware under a constant input stimulus.

\begin{figure}[tbp]
    \centering

    % (a) Tonic Bursting
    \begin{minipage}[t]{0.48\linewidth}
        \centering
        \includegraphics[width=\linewidth]{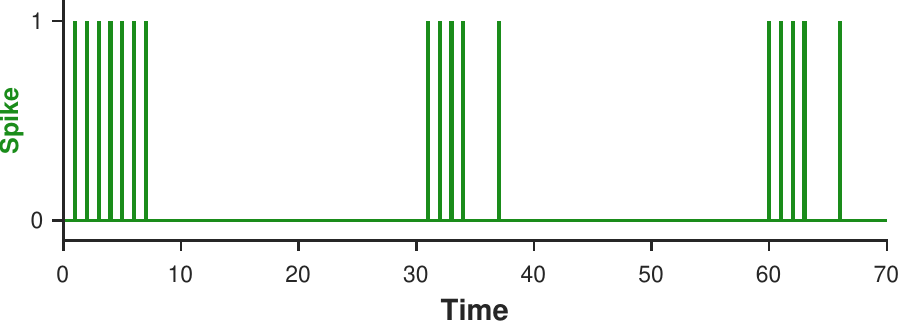}
        
        \scriptsize (a) Tonic Bursting
    \end{minipage}
    \hfill
    % (b) Phasic Bursting
    \begin{minipage}[t]{0.48\linewidth}
        \centering
        \includegraphics[width=\linewidth]{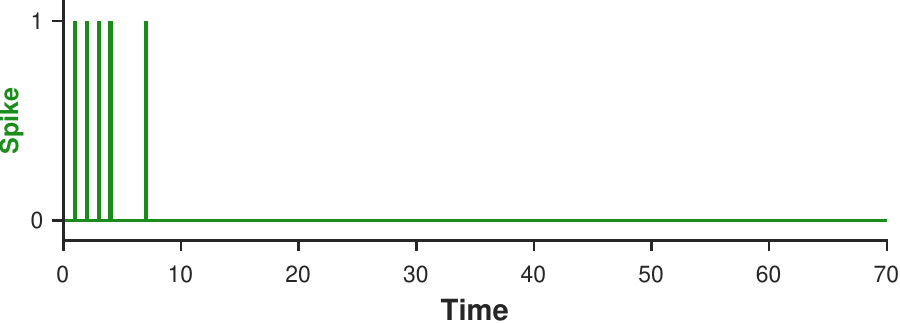}
        
        \scriptsize (b) Phasic Bursting
    \end{minipage}

    \caption{\textbf{Software emulation of multi-pattern neural dynamics}
    (a) Simulated \textbf{tonic-bursting} response showing repetitive clusters 
    of high-frequency spikes interspersed with silent refractory periods under 
    a constant input stimulus ($I_{in} = 1.0$). 
    (b) Simulated \textbf{phasic-bursting} response exhibiting a transient 
    high-frequency burst at stimulus onset followed by sustained quiescence.}
    
    \label{fig:AdEx_Response}
\end{figure}

We assess the functional significance of rich neuronal dynamics beyond their hardware realization by algorithmically simulating tonic-bursting and phasic-bursting responses to serve as software proxies for our device. As illustrated in Fig.~\ref{fig:AdEx_Response}, the algorithmically modeled neurons successfully reproduce the distinct bursting behaviors generated by the hardware under a constant input stimulus.

We incorporated these neuron models into spiking neural networks (SNNs) and evaluated them across four benchmark datasets: MNIST~\cite{lecun1998gradient}, Fashion-MNIST~\cite{xiao2017fashion}, Breast Cancer~\cite{uci_breast_cancer}, and IRIS~\cite{iris_53}. The performance of these networks was then compared against an SNN of identical architecture utilizing conventional leaky integrate-and-fire (LIF) neurons as a non-bursting baseline (detailed emulation and network training methodologies are provided in the Methods section).

As outlined in Table  \ref{tab:AdEx_Result}, both bursting configurations achieved a statistically significant reduction in spike activity across all four datasets ($p < 0.001$, Welch’s t-test) while maintaining classification accuracy comparable to the LIF baseline. For instance, on the MNIST dataset, the standard LIF baseline achieved 97.80\% accuracy but required an average of 9,326 spikes per image. In contrast, the phasic bursting mode maintained an accuracy of 97.59\% while reducing the average spike count to only 3,449 (a 63\% decrease in spike activity). A similar algorithmic advantage was observed with the Fashion-MNIST dataset, where the phasic bursting mode yielded a slight accuracy improvement over the LIF baseline (86.76\% versus 85.21\%) while decreasing the average spike count by nearly an order of magnitude, dropping from 42,861 to 4,916 spikes. This trend of preserved accuracy and enhanced sparsity extended to the smaller-scale tabular datasets, with the phasic bursting mode reducing total spike activity by approximately 57\% and 80\% on the Breast Cancer and IRIS datasets, respectively.

% As outlined in Table \ref{tab:AdEx_Result}, both bursting configurations substantially reduced spike activity across all four datasets while maintaining classification accuracy comparable to the LIF baseline.

\begin{table*}[tbp]
\centering
\caption{\textbf{Comparison of classification accuracy and spike activity across neuron models.} SNNs utilizing bursting dynamics (Fig.~\ref{fig:AdEx_Response}) drastically reduce the network spike count required per inference compared to the LIF baseline, without suffering notable degradation in task performance. ($*\ p < 0.001$ versus LIF baseline, Welch's t-test.) }
\label{tab:AdEx_Result}
\resizebox{\columnwidth}{!}{%
\begin{tabular}{@{}ccccccc@{}}
\toprule
\textbf{}              & \multicolumn{3}{c}{\textbf{Classification Accuracy (\%)}} & \multicolumn{3}{c}{\textbf{Spikes per Inference (Mean $\pm$ SD)}} \\ \cmidrule(lr){2-4} \cmidrule(lr){5-7}
\textbf{Dataset} &
  \multicolumn{1}{c}{\textbf{LIF}} &
  \multicolumn{1}{c}{\textbf{Tonic Burst}} &
  \multicolumn{1}{c}{\textbf{Phasic Burst}} &
  \multicolumn{1}{c}{\textbf{LIF}} &
  \multicolumn{1}{c}{\textbf{Tonic Burst}} &
  \multicolumn{1}{c}{\textbf{Phasic Burst}} \\ \midrule
\textbf{MNIST}         & 97.80             & 97.59             & 97.59             & 9,326 $\pm$3,581& 5,177 $\pm$568*      & \textbf{3,449 $\pm$147}*    \\
\textbf{Fashion-MNIST} & 85.21             & 85.73             & 86.76             & 42,861$ \pm$12,948     & 13,707 $\pm$3,708*  & \textbf{4,916 $\pm$647}*    \\
\textbf{Breast Cancer} & 97.39             & 97.39             & 97.39             & 765 $\pm$423        & 444 $\pm$81*       & \textbf{328 $\pm$16}*     \\
\textbf{IRIS}          & 96.77             & 96.77             & 96.77             & 2,123 $\pm$332      & 750 $\pm$56*        & \textbf{431 $\pm$8}*      \\ \bottomrule
\end{tabular}%
}
\end{table*}

This reduction in spike activity complements the low per-spike energy of the proposed hardware. Because the V-MTJ/CMOS neuron supports these bursting dynamics at an energy cost as low as 8.82 fJ per spike (e.g., during tonic bursting), reducing the number of spike events provides a potential pathway toward lower spike-mediated energy consumption. More broadly, co-designing the hardware to support diverse firing dynamics can reduce both neuronal switching activity and spike-mediated communication in densely interconnected neuromorphic systems.

%---------------------------------------------------------------------------
% End of Software Section 
%---------------------------------------------------------------------------

\subsubsection*{Conclusion}

This work presented a reconfigurable neuromorphic neuron that combines measured VCMA-MTJ device dynamics with CMOS circuitry to emulate diverse biologically inspired firing behaviors. By leveraging the measured V-MTJ state dynamics and a compact CMOS membrane and recovery circuit, the proposed architecture reproduces diverse biologically inspired firing behaviors, including tonic spiking, phasic spiking, tonic bursting, phasic bursting, and spike latency, without requiring device resizing or dedicated neuron circuits for individual firing patterns. Circuit-level validation in the commercial GlobalFoundries 22nm FD-SOI technology demonstrates the feasibility of integrating measured V-MTJ device dynamics with advanced CMOS circuitry to realize compact and energy-efficient neuromorphic neurons. Furthermore, algorithmic simulations confirm that deploying these diverse dynamics within spiking neural networks substantially reduces total spike activity. This proves that multi-pattern neuronal behaviors can serve as a direct mechanism for achieving highly sparse, computationally efficient inference without compromising classification accuracy. The proposed device-circuit co-design establishes a scalable framework for reconfigurable spintronic neurons capable of emulating diverse biologically inspired firing behaviors, providing a promising foundation for next-generation brain-inspired computing systems.

\section*{Methods}

\subsection*{MTJ Fabrication and Characterization}
The V-MTJ devices used in this work were fabricated on thermally oxidized silicon substrates using ultra-high-vacuum magnetron sputtering. The multilayer stack consisted of a synthetic antiferromagnetic reference layer, W (0.25 nm), CoFeB (0.8 nm), MgO (1.5 nm), CoFeB (1.6 nm), and a Mo (5 nm) capping layer \cite{athas2025statistical}. Following deposition, the films were annealed to crystallize the CoFeB/MgO interface and establish perpendicular magnetic anisotropy. Circular magnetic tunnel junctions were subsequently patterned using electron-beam lithography and ion milling with diameters of 50 nm, followed by dielectric encapsulation and top-electrode metallization to complete the device fabrication.

Electrical characterization was performed at room temperature using a custom high-speed measurement platform. A constant read voltage was applied across the V-MTJ while the device resistance was continuously monitored to capture stochastic transitions between the P and AP magnetic states. Representative resistance-time traces exhibiting random telegraph noise were recorded under different applied voltage and field conditions, and the dwell times of the P and AP states were extracted statistically from the measured switching events. The average dwell times and state occupation probabilities were subsequently used to characterize the voltage-dependent switching dynamics of the V-MTJ and serve as the experimental foundation for the neuron model presented in this work.

\subsection*{Spiking Neural Network Architecture and Optimization}

To evaluate the computational implications of the proposed hardware, the observed multi-pattern firing dynamics were algorithmically emulated within fully connected spiking neural networks (SNNs). We have modeled these Izhikevich-inspired temporal behaviors using the Adaptive Exponential Integrate-and-Fire (AdEx) framework, alongside a standard Leaky Integrate-and-Fire (LIF) baseline to serve as continuous-time software proxies~\cite{brette2005adaptive,naud2008firing}. The SNN architecture comprised a single hidden layer containing 3,000 spiking neurons, with the output layer dimensionality matching the target classes of the respective datasets. The temporal dynamics of all input samples were discretized using a forward Euler integration scheme ($dt=1.0$) and evaluated over a sequence of 40 discrete time steps. 

All models were implemented using the PyTorch machine learning framework \cite{paszke2019pytorch} in conjunction with the snnTorch library \cite{eshraghian2023training}. Network parameters were optimized using the Adam optimizer~\cite{kingma2014adam} with a learning rate of $5 \times 10^{-5}$ and momentum parameters $\beta_1 = 0.9$ and $\beta_2 = 0.999$. The objective function was computed using standard Cross-Entropy Loss applied to the temporal accumulation of the output layer spikes (integrated logits). To perform gradient descent through the non-differentiable step function of the spike generation mechanism, network optimization leveraged Backpropagation Through Time (BPTT)~\cite{neftci2019surrogate} coupled with a surrogate gradient approach. Specifically, we utilized the fast sigmoid surrogate derivative provided by snnTorch~\cite{eshraghian2023training}, defined by a steepness slope parameter of 40.0.

\section*{Acknowledgments}

The work at the University of Wisconsin–Madison was supported in part by the U.S. Department of Energy under Award No. DE-SC0026035 and U.S. National Science Foundation under the award number CCF-2319617. The work at George Mason University was supported by the U.S. Department of Energy under Award No. DE-SC0026260 and the National Science Foundation under Award No. 2539714. The work at Northwestern University was supported by the U.S. Department of Energy under Award No. DE-SC0026325. The authors used AI for proofreading and paraphrasing, not for scientific content generation.

\section*{Author contributions}
K.A. conceptualized the study, designed and verified the circuits, and developed the device–circuit co-design framework. J.A. conducted the experimental design, data collection, and analysis for the characterization and switching behaviour of the magnetic tunnel junctions (MTJs). A.F. assisted with data collection and analysis for the characterization and switching behaviour of the MTJ. H.P. developed the software emulation, performed the SNN simulations and result analysis, and drafted the corresponding algorithmic sections. A.J. also conceptualized the study and supervised the overall research. G.A., M.P., and P.A. supervised the neuroscience, device and algorithm aspect of the research respectively. All authors contributed to reviewing and editing the manuscript.

\section*{Competing interests}
The authors declare no competing interests.

\label{subsec2}

\vspace{1em}

\FloatBarrier


\bibliography{sn-bibliography}% common bib file

\end{document}